\documentclass[twocolumn,showpacs,preprintnumbers,amsmath,amssymb]{revtex4}

\usepackage{graphicx}
\usepackage{dcolumn}
\usepackage{bm}

\begin{document}

\title{Dissociative recombination of BeH$^+$}

\author{J. B. Roos, M. Larsson, and \AA. Larson\footnote{Corresponding author; e-mail: aasal@physto.se}}
\affiliation{Dept. of Physics, Stockholm University, AlbaNova University Center
 S-106 91 Stockholm, Sweden}
\author{A. E. Orel}
\affiliation{Dept. of Applied Science, University of California, Davis, CA 95616, USA}

\date{\today}

\begin{abstract}
The cross section for dissociative recombination of BeH$^+$ is calculated by solution of the 
time-dependent Schr\"{o}dinger equation in the local complex potential approximation. 
The effects of couplings between resonant states and the Rydberg states converging to the ground state of the ion are studied.
The relevant potentials, couplings and autoionization widths are extracted using \textit{ab initio} electron
scattering and structure calculations, followed by a diabatization procedure. The calculated cross sections
shows a sizable magnitude at low energy, followed by a high-energy peak centered around 1 eV. The electronic couplings between the neutral states induce oscillations in the cross section. Analytical forms for the cross sections at low collision energies are given.
\end{abstract}

\maketitle
\date{\today}

\section{\label{sec:intro}Introduction}

The ITER fusion reactor planned for Cadarache, France, is the next 
major step in fusion research. It is presently scheduled to be taken
into operation in 2016. A critical choice in the construction of ITER
is the plasma facing material. There is not a single choice which will
fulfil all possible criteria, and research and development is required
in order to make an optimal choice. Much efforts have been devoted
to obtain reaction cross sections for atoms and molecules present in the
edge plasma and the divertor region, not least through 
the Coordinated Research Projects organised by IAEA~\cite{clark06}.
Beryllium is a primary choice for the
first wall because of its low erosion and atomic number. Even with a low 
degree of erosion, however, beryllium will enter the plasma and it is 
therefore of importance to include surface and volume processes for beryllium
in various forms in modelling the edge plasma. 
Beryllium-containing hydrides will form in the edge plasma, and these molecules
will be ionized by electrons. The question posed in this paper is how effectively
the BeH$^+$ ions and its isotopologs are destroyed by electrons. The dominant process
destroying BeH$^+$ in the edge plasma is dissociative recombination. This process has never,
as far as we know, been studied experimentally, because of the toxicity of beryllium.
Thus, there is a need for theoretical support to determine the cross section for this
reaction. In this paper we present the
first calculation of the dissociative recombination of BeH$^+$.

The relevant potential curves and autoionization widths are determined using \textit{ab initio}
electron scattering and structure calculations. This is followed by a diabatization procedure, where
the electronic couplings between the neutral states can be determined. 
This is depicted in section~\ref{sec:pec}. The nuclear dynamics is then studied using a wave
packet propagation method described in section~\ref{sec:nuc}. 
The results are presented in section \ref{sec:res}. Atomic units are used throughout the paper unless otherwise stated.

\section{Calculation of relevant electronic states\label{sec:pec}}

The potential energy curves of the electronic states relevant for dissociative 
recombination of BeH$^+$ are determined
by combining structure calculations with electron scattering calculations.
For the structure calculations, the Multi-Reference Configuration Interaction (MRCI)
technique is used to determine the neutral electronically bound 
adiabatic states situated below the potential
energy curve of the ground state ($X^1\Sigma^+$) of the BeH$^+$ ion.

In a quasidiabatic representation, some of these neutral states
will cross the ionic ground state and become resonant states, i.e., they will couple to the ionization
continuum and have a finite probability for autoionization. In order to determine the energies and autoionization
widths of the resonant states, the Complex Kohn Variational method~\cite{rescigno95} is used.
In order to obtain the quasidiabatic potentials, the scattering data is combined with a diabatization
procedure of the electronically bound states. 

Finally, the calculated potential energy curves, autoionization widths
and electronic couplings between the neutral states can be used in
order to carry out calculations on the nuclear dynamics.

\subsection{\label{sec:struc}Structure calculations}

The adiabatic potential energy curve of the BeH$^+$ ground state ($X^1\Sigma^+$) as well as several excited 
states of BeH of $^2\Sigma^+$, $^2\Pi$, and $^2\Delta$ symmetries 
are calculated using the MRCI technique.

For both the structure and scattering calculations, natural orbitals are used. The natural orbitals
are determined using a MRCI calculations on the BeH$^+$ ground state. In this calculation,
the reference space consists of the $1\sigma,2\sigma,3\sigma,1\pi,4\sigma$ orbitals and single and double excitations
from the reference configurations into the virtual orbitals are included.
The natural orbitals are calculated using a basis set for the H atom of ($4s$,$1p$) contracted to
[$3s$,$1p$]~\cite{dunning70}, while for Be, a basis set of ($10s$,$6p$,$1d$) contracted into [$3s$,$3p$,$1d$] is used~\cite{dunning77,feller96}. The natural orbitals are then further expanded with diffuse orbitals in order to accurately describe the Rydberg character of some of the electronic states. The H-orbitals are augmented with ($2s$,$2p$,$1d$) orbitals, and the Be-orbitals
with ($4s$,$1p$,$1d$). 

The MRCI calculations on the ionic and neutral excited electronic states are carried out using
a reference space consisting of the $1\sigma,2\sigma,3\sigma,1\pi,4\sigma,5\sigma$ orbitals. 
These natural orbitals all have an occupation number greater than approximatly 0.002.

In addition single excitations out of the reference space are included. The calculations are carried out in $C_{2v}$ symmetry. For the neutral system,
the MRCI calculations consist of about 5500 configurations in $A_1$ symmetry, 3700 in $A_2$ and 4600 in $B_1$.
For all symmetries, the 30 lowest roots are calculated. The potential energy curves are calculated for internuclear distances ranging from 1 a$_0$ to 10 a$_0$. 

\begin{figure}[h]
\rotatebox{0.0}{\includegraphics[width=0.9\columnwidth]{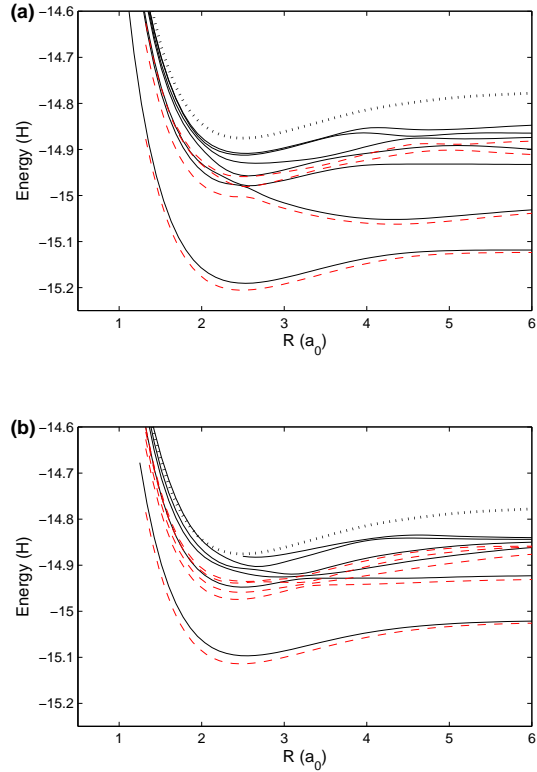}}
\caption{\label{fig1} Adiabatic potential energy curves of BeH of (a) $^2\Sigma^+$ symmetry, and (b) $^2\Pi$ symmetry. The black dotted curve is the ground state ($X^1\Sigma^+$) of the BeH$^+$ ion. The black full curves are the potentials
calculated here, while the dashed (red online) curves are potentials calculated by Pitarch-Ruiz \textit{et al.}~\cite{pitarch-ruiz08}} 
\end{figure}

The adiabatic potential energy curves situated below the ionic
ground state potential are displayed in Fig.~\ref{fig1}.
In (a) we show the electronic states of BeH of $^2\Sigma^+$ symmetry and in (b) we show the corresponding
states of $^2\Pi$ symmetry. The dotted curve is the potential energy curve of the ionic ground state. The dashed (red online) curves included in the figure are the very accurate potentials of BeH recently calculated by Pitarch-Ruiz \textit{et al.}~\cite{pitarch-ruiz08} using a full configuration interaction calculation with a large basis set. As can be seen in the figure, the potential energy curves calculated by Pitarch-Ruiz \textit{et al.} are lower in absolute energy. However, our potentials show a similar dependence as a function of the internuclear distance. In the present calculation, our structure calculations are limited by the scattering calculations we perform in the next step (see section~\ref{sec:scat}). In order to describe the resonant and electronically bound states at the same level of theory, it is important to use the same MRCI wave function for the target ion in both the structure and the scattering calculations. In Fig.~\ref{fig1}, we do not display any $^2\Delta$ states. Our basis set is not good enough to represent any Rydberg $^2\Delta$ states. We only obtain one resonant state of this symmetry. The higher excited states of $^2\Delta$ symmetry have threshold energies above $1.05$ eV relative to the ground vibrational level of the ion and these states will not be important for the total cross section of dissociative recombination at low collision energies.	

\subsection{\label{sec:scat}Electron scattering calculations}

By using the Complex Kohn variational method~\cite{rescigno95}, the energy positions and autoionization widths of the resonant states are determined. As a target wave function $\bf \Phi_i(\bf{r}_1,...,\bf{r}_N)$, the same MRCI wave function for the ion as used in the structure calculations is applied. In the Complex Kohn variational method, the $N+1$ electron trial wave function is written as a
\begin{equation}
{\bf \Psi}=\sum_{i}A{\lbrack{\bf \Phi}_{i}({\bf r}_{1}...{\bf r}_{N})F_{i}({\bf r}_{N+1})\rbrack}+\sum_{\mu}d_{\mu}{\bf \Theta}_{\mu}({\bf r}_{1}...{\bf r}_{N+1}).
\end{equation} 
The first sum is denoted as the $P$-space portion of the wave
function and runs over the energetically open target states. In the case of
BeH, only one channel was open.
 The function $F_{i}({\bf r}_{N+1})$ is the one-electron wave function describing the
scattered electron, and $A$ is an anti-symmetrization operator for the electronic coordinates.
The second term, denoted as the $Q$-space portion of the wave function, contains the functions 
${\bf \Theta}_{\mu}$, which are square-integrable $N+1$ configuration state functions (CSFs) which are used to describe
short-range correlations and the effects of closed channels. The advantage of using natural orbitals is
that the orbital space used to generate these states is kept manageable small.
The one-electron scattering wave function $F_i$ is in the case of electron-ion scattering, further expanded as
\begin{eqnarray}
F_{i}({\bf r})&= & \sum_{i}c_{j}^{i}\phi_j({\bf r}) + \\ \nonumber & & \left[f_{l}^{-}(k_{i}r)\delta_{ll_{0}}\delta_{mm_{0}}+T_{ll_0mm_0}^if_{l}^{+}(k_{i}r)\right]Y_{lm}(\hat{{\bf r}})/r.
\end{eqnarray}
Here $\phi_j({\bf r})$ are a set of square-integrable functions, $f_{l}^{\pm}$ are the outgoing- and incoming Coulomb functions and $Y_{lm}$ are spherical Harmonics. Angular momenta up to $l=6$ and $|m|=4$ are included in the calculation.

By inserting the trial wave function into the Complex Kohn Functional~\cite{rescigno95}, the unknown coefficients in the trial wave function can be optimized. Then also the $T$-matrix ($T_{ll_0mm_0}^i$) for elastic scattering is obtained and by fitting the eigenphase sum of the $T$-matrix to a Breit-Wigner form~\cite{geltman97}, the energy positions and autoionization widths of the resonant states can be calculated. These electron-scattering calculations are carried out for a fixed geometry of the target ion. For BeH, the five lowest resonant states of $^2\Sigma^+$ and $^2\Pi$ symmetries as well as the lowest resonant state of $^2\Delta$ symmetry are calculated. The electron scattering calculations are carried out for $1.0$ a$_0$ $\leq R \leq 3.0$ a$_0$ with steps of $0.25$ a$_0$.

\subsection{\label{sec:diab}Diabatization}

For internuclear distances smaller than $5.0$ a$_0$, the electronic ground state of BeH$^+$ has the dominant configuration $(1\sigma)^2(2\sigma)^2$. All Rydberg states converging to this ionic core have this configuration plus an extra electron in an outer, Rydberg-like orbital. The resonant states are in general Rydberg states converging to excited ionic cores. For these states the $(2\sigma)$ orbital is typically singly occupied. It is therefore relative easy to use the configuration interaction coefficients of the MRCI wave functions to follow the resonant states when they cross the Rydberg manifold situated below the ionic ground state potential. This is done in order to obtain an ``estimate" of the quasidiabatic potential energy curves. 

It should be mentioned that for $R \geq 5.0$ a$_0$, the ionic ground state changes dominant configuration to $(1\sigma)^2(2\sigma)^1(3\sigma)^1$. For these large
internuclear distances, the Rydberg states will then have similar configurations as the resonant states and the quasidiabatic potentials are simply estimated by the shape of the adiabatic potentials. 

\begin{figure}[h]
\rotatebox{0.0}{\includegraphics[width=0.9	\columnwidth]{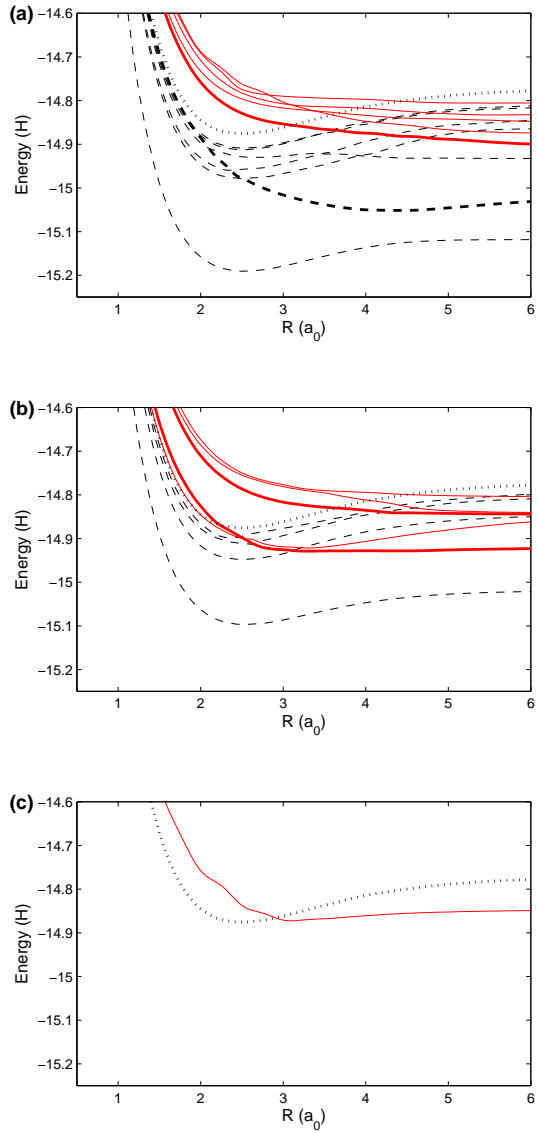}}
\caption{\label{fig2} Quasidiabatic potentials of BeH of (a) $^2\Sigma^+$, (b) $^2\Pi$ and (c) $^2\Delta$ symmetries.
The solid (red online) curves are the resonant states while the electronically bound states are dashed (black) curves.
Also the electronic ground state of the BeH$^+$ is displayed with the dotted curve.} 
\end{figure}

This technique, however, will only provide us with diagonal element of the diabatic potential matrix and not the electronic couplings between the diabatic states. In order to obtain these couplings a diabatization procedure is applied. This diabatization procedure was very recently developed by us for determining the quasidiabatic potential matrix for the HF molecule~\cite{roos09}.
There is a unitary transformation matrix ${\bf T}$ that will transform the adiabatic potential matrix [with matrix elements
$V_{ij}^a=V_i^a(R)\delta_{ij}$] into the diabatic potential matrix
\begin{equation}
\label{eq:vd}
{\bf V}_d={\bf T}{\bf V}_a {\bf T}^{-1}.
\end{equation}
In this diabatization procedure, we assume that the transformation matrix can be written as a product of matrices that describe successive 2x2 rotations among the adiabatic states ${\bf T}=\ldots{\bf T}_2{\bf T}_1$. The matrices ${\bf T}_i$ are of the form (in the case of a rotation among state $1$ and $2$)
\begin{equation}
{\bf T}_i=\left(
\begin{array}{ccccc}
	\cos(\gamma_i) & -\sin(\gamma_i) & 0 & 0 & \ldots \\
	\sin(\gamma_i) & \cos(\gamma_i) & 0 & 0 & \ldots \\
	0 & 0 & 1 & 0 & \ldots \\
	0 & 0 & 0 & 1 & \ldots \\
	\vdots & \vdots & \vdots & \vdots & \ddots \\
\end{array}
\right),
\end{equation}
where we assume that the rotational angles have the analytical form
\begin{equation}
\gamma_i=\frac{\pi}{4}\left[1+\tanh\left(\alpha_i(R-R_i)\right)\right].
\end{equation}
We then perform an optimization procedure, where the unknown parameters $\alpha_i$ and $R_i$
of the rotational angles are determined in order to optimize the agreement between the diagonal
elements of the diabatic potential matrix in Eq.~(\ref{eq:vd}) and the estimated quasidiabatic potentials
obtained using the CI-coefficients as described above. When the transformation matrix is optimized, not
only the diagonal diabatic potential curves are obtained, but also the electronic couplings between the 
neutral states.
In Fig.~\ref{fig2} we show the diabatic potential curves of BeH obtained using this procedure. 

\begin{figure}[h]
\rotatebox{0.0}{\includegraphics[width=1.0\columnwidth]{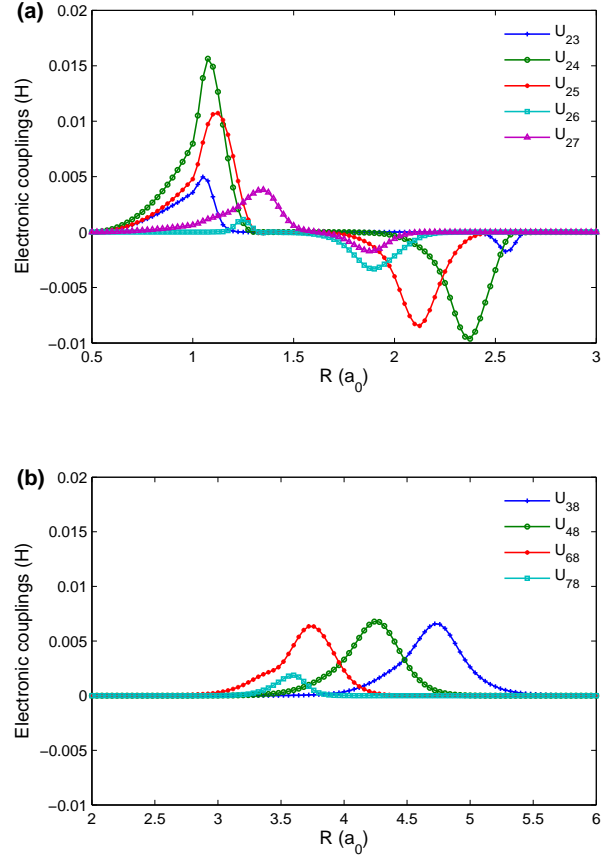}}
\caption{\label{fig3} Some of the electronic couplings between diabatic states of $^2\Sigma^+$ symmetry.} 
\end{figure}

The resonant states situated above the ionic ground state are calculated using the electron scattering calculations. These resonant states
will then cross some of the Rydberg states situated below the ionic potential. Where these curves cross, they will couple by electronic couplings. In Fig.~\ref{fig3}, we show examples of the electronic coupling elements between the $^2\Sigma^+$ neutral states marked with the thicker curves in Fig.~\ref{fig2} with the other states of the same symmetry. We label the diabatic electronic states from 1 to $n$ in energy order from lowest to highest
at the internuclear distance 1.0 a$_0$.
It should be noted that the electronic couplings are localized to the regions of avoided crossings.
The couplings between the electronically bound state 2 and the higher states [see Fig~\ref{fig3} (a)] have the double peak structure
since the compact state 2 crosses several of the Rydberg states twice. This state does not, however, cross the ionic ground state.

\begin{figure}[h]
\rotatebox{0.0}{\includegraphics[width=1.0\columnwidth]{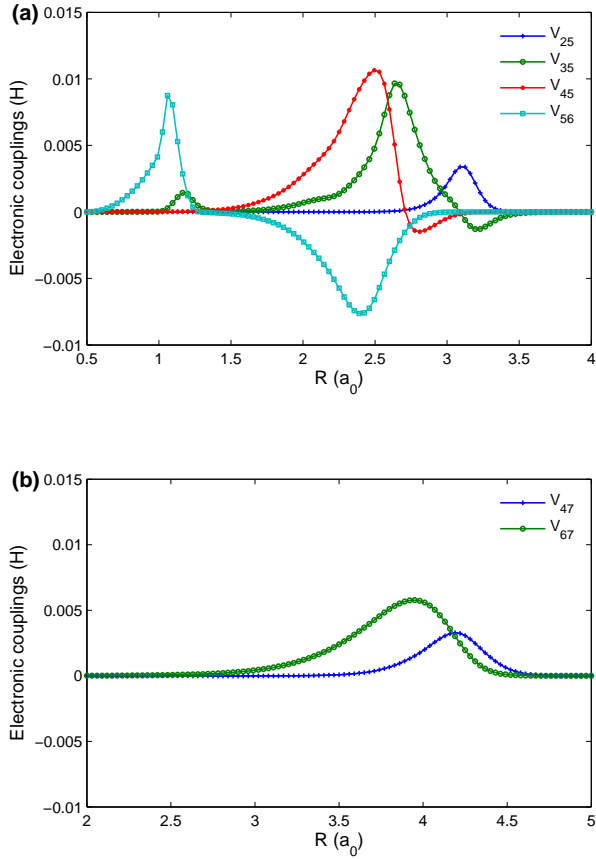}}
\caption{\label{fig4} Some of the electronic couplings between diabatic states of $^2\Pi$ symmetry.} 
\end{figure}

Similar couplings, but now for the $^2\Pi$ manifold are displayed in Fig~\ref{fig4}.

\section{\label{sec:nuc}Nuclear dynamics}

The cross section for dissociative recombination of BeH$^+$ is calculated by propagating wave packets on the neutral
states. The electron capture will induce the initial condition for the wave packets on the resonant state $i$~\cite{mccurdy83}
\begin{equation}
\label{eq:init}
\Psi_i(t=0,R)=\sqrt{\frac{\Gamma_i(R)}{2\pi}}\chi_{v}(R).
\end{equation}
Here $\Gamma_i$ is the autoionzation width for the resonant state and $\chi_{v}$ is the vibrational wave function of the ion. The rotational motion of the molecule is neglected.

The nuclear dynamics is studied by numerically integrating the time-dependent Schr\"{o}dinger equation
\begin{equation}
\label{eq:se}
i\frac{\partial}{\partial t}{\bf \Psi}(t,R)=-\frac{1}{2\mu}{\bf I}\frac{\partial^2}{\partial R^2}{\bf \Psi}(t,R) + 
{\bf V}_{d}{\bf \Psi}(t,R)
\end{equation}
using a Cranck-Nicholson propagator~\cite{goldberg67}. In the present study, the nuclear dynamics is studied using both coupled and uncoupled electronic states. For the uncoupled states, ${\bf V}_d$ is a diagonal matrix with the quasidiabatic potentials of the resonant states on the diagonal. Autoionization is included by letting the resonant states become complex when the resonant state potential is situated above the ionic ground state
\begin{equation}
\tilde{V}_i^d(R)=V_i^d(R)-i\frac{\Gamma_i(R)}{2}.
\end{equation}
This is the so-called local approximation for treating autoionization~\cite{mccurdy83}. The validity of this approximation is examined using a
non-local expression for autoionization~\cite{giusti83} and we found that even for the two lowest resonant states of BeH of $^2\Pi$ symmetry, the local approximation is valid.
When the electronic couplings are included, the matrix ${\bf V}_d$ in eq. (\ref{eq:se}) now contains both
the resonant and electronic bound state potentials on the diagonal as well as the off-diagonal coupling elements.
The electron capture directly into the Rydberg states as well as autoionization out of the Rydberg states are in the
present study neglected. The neglect of the so-called indirect mechnism is valid when there is a strong direct mechnism~\cite{bates91,florescu03,ngassam06}, i.e., when there
are resonant states with potentials crossing the ion close to its minimum. The indirect mechnism will only affect the cross section at very low
collision energies and create sharp structures in the cross section.

The wave packets are propagated out into the asymptotic region, where the cross section for dissociative recombination is calculated~\cite{mccurdy83,kulander78} by projecting the asymptotic wave packets onto the energy-normalized eigenstates of the fragments
\begin{equation}
\sigma_i(E)=\frac{2\pi^3}{E}g\left|\langle\Phi^i_E(R)|\Psi_i(t_{\infty},R)\rangle\right|^2.
\end{equation}
Here $g$ is the ratio of multiplicity between the the resonant state and ionization continuum. The total cross section for dissociative recombination is obtained by symmarizing all partial cross sections $\sigma(E)=\sum_i{\sigma_i(E)}$.

In the present calculations, the wave packets are propagated on a grid ranging from $0.5$ a$_0$ to $300$ a$_{0}$ with
$dR=0.01$ a$_0$. The wave packets are propagated with time-steps of $dt=0.1$ a.u. For the uncoupled potentials, the wave packets are propagated until $t_{\infty}=1000$ a.u., while $t_{\infty}=4000$ a.u. is needed for the coupled system.

\section{\label{sec:res}Results and discussion}

As mentioned above two models are applied in the study of dissociative recombination of BeH$^+$. In the first model, only the dynamics on the uncoupled resonant states is considered, while in the second model the propagation of the wave packets on the coupled potentials is explored. In both models, autoionization from the resonant states is included using local complex potentials.

\subsection{Uncoupled states}

\begin{figure}[h]
\rotatebox{0.0}{\includegraphics[width=0.85\columnwidth]{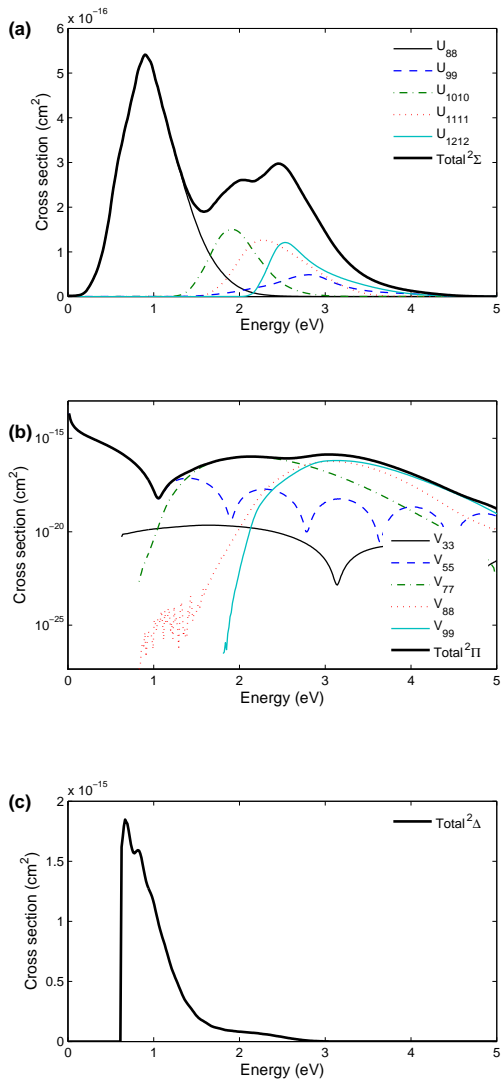}}
\caption{\label{fig5} Cross section for dissociative recombination of BeH$^+$, calculated by propagating wave packets on uncoupled resonant states of (a) $^2\Sigma^+$ symmetry, (b) $^2\Pi$ symmetry (using a linear-log scale) and (c) $^2\Delta$ symmetry. The thin curves are the cross sections originating from each resonant state, while the thick curve is the total cross section from the states of a given symmetry.} 
\end{figure}

As described above, five resonant states of $^2\Sigma^+$ symmetry are included in the model.
Only the two lowest resonant states (here labelled with $U_{88}$ and $U_{99}$) are associated with asymptotic 
limits that are energetically open at zero collision energy. The following $^2\Sigma^+$ states open up for dissociation at energies of 0.56 eV ($U_{1010}$), 0.87 eV ($U_{1111}$) and 1.59 eV ($U_{1212}$) relative to the ground vibrational level of BeH$^+$. In the following the cross section for electron recombination with BeH$^+$ in its ground vibrational state is calculated, that is, the vibrational wave function of the $v=0$ level of BeH$^+$ is used in Eq. (\ref{eq:init}).
In Fig.~\ref{fig5} (a), we display the cross section for dissociative recombination, obtained when propagating the wave
packets on the uncoupled $^2\Sigma^+$ states.
As can be seen in the figure, the cross sections originating from all $^2\Sigma^+$ states show a smooth onset. This reflects the fact that the resonant states for this symmetry do not cross the ionic ground state potential close to the equilibrium separation of the ion. The energy-dependence at threshold is therefore determined by the capture probability which is related to the Franck-Condon factors between the initial vibrational wave function of the ion and the continuum wave function of the resonant states. 

The situation is different for the resonant states of $^2\Pi$ symmetry, where two states (labelled with $V_{33}$ and $V_{55}$) cross the ion potential at smaller distances than the equilibrium distance. The first resonant state ($V_{33}$) has a threshold of about 0.63 eV, while the second state ($V_{55}$) is energetically open for dissociation at zero collision energy. The cross section from this state shows the typical $1/E$ dependence~\cite{wigner48} of the cross section at low energies. The remaining states of $^2\Pi$ symmetry have threshold energies of about 0.83 eV (for $V_{77}$ and $V_{88}$) and 1.81 eV for $V_{99}$. As can be seen in Fig.~\ref{fig5} (b), where the cross section from the $^2\Pi$ states are plotted using a linear-log scale, it is the low energy tail of the cross section from the second resonant state ($V_{55}$) that dominates the low-energy cross section for this symmetry. 

Finally, in Fig.~\ref{fig5} (c), the cross section from the $^2\Delta$ resonant state is displayed. This state opens up for dissociation at 0.63 eV and the cross section shows a sharp threshold at this energy.

\begin{figure}[h]
\rotatebox{0.0}{\includegraphics[width=0.9\columnwidth]{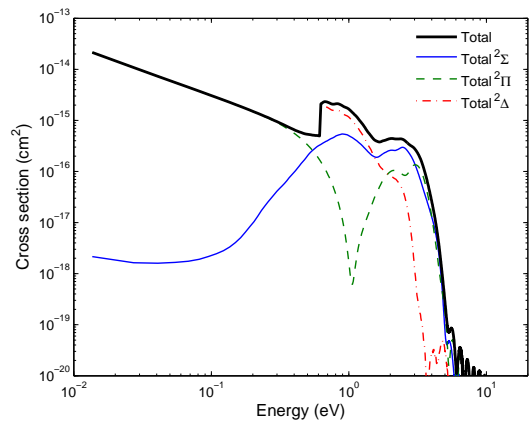}}
\caption{\label{fig6} Total cross section for dissociative recombination of BeH$^+$, calculated using uncoupled
resonant states.} 
\end{figure}

In Fig.~\ref{fig6}, the total cross section using a log-log scale calculated with uncoupled resonant states is shown. In the figure, also the contribution to the cross section from states of the different symmetries are displayed. As can be seen, at low energy the cross section is dominated by resonant states of $^2\Pi$ symmetry. The sharp threshold around 0.63 eV, asising from the $^2\Delta$ resonant state can be seen in the total cross section.

\subsection{Coupled states}

As mentioned above, several resonant states of BeH are not energetically open for dissociation at zero collision energy.
We therefore wish to examine a model where the electronic couplings between the resonant and electronically bound states are included. We have previously seen for other systems such as HF$^+$~\cite{roos09,roos08} that the inclusion of electronic couplings between the neutral states may open up pathways to dissociation. Flux that in the uncoupled model is trapped in bound potentials will transfer to lower states that are open for dissociation.

\begin{figure}[h]
\rotatebox{0.0}{\includegraphics[width=0.9\columnwidth]{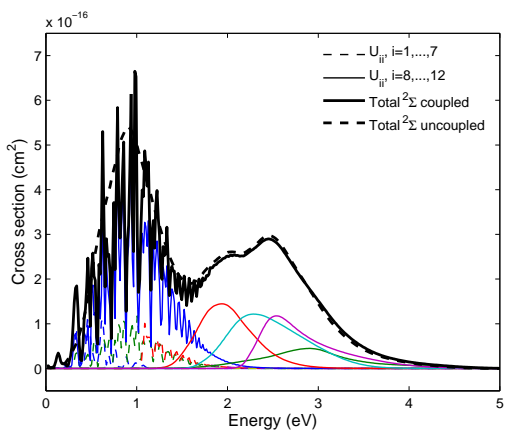}}
\caption{\label{fig7} Cross section calculated using the coupled states of BeH of $^2\Sigma^+$ symmetry.
The thin solid curves are the cross sections from the resonant states, while the dashed curves are cross sections from the electronically bound states. The total cross section originating from the coupled $^2\Sigma^+$ state is shown with the thick black curve. Also the cross section calculated using uncoupled $^2\Sigma^+$ states is displayed with a thick dashed black curve.} 
\end{figure}

As described above in section~\ref{sec:diab}, the neutral states are diabatized and the electronic couplings are determined.
For the states of $^2\Sigma^+$ symmetry, seven states are diabatized, i.e., the lowest resonant state and six electronically bound states. In a quasidiabatic representation, there is one compact electronically bound state that will cross several of the Rydberg states situated below the ionic ground state. However, this state will never cross the ion potential and become a resonant state. The ground state of BeH, $X^2\Sigma^+$, is assumed not to couple to the higher electronic states. 
Due to numerical limitations in the optimization procedure, the electronic couplings to the higher resonant states of this symmetry are not calculated. The wave packets are then propagated using the full diabatic potential matrix. In Fig.~\ref{fig7}, the cross section, calculated using the coupled $^2\Sigma^+$ states is shown. In the figure, also the cross section calculated using uncoupled states are dispalyed. 

The inclusion of the electronic couplings will not dramatically change the magnitude of the cross section. Sharp structure
is created both in the partial cross sections as well as the total cross section. This structure can be interpreted as Feshbach resonances created when the states open for dissociation are coupled to the nuclear bound states. Structures in the cross section is then formed when the wave packets can couple from a resonant state and form a vibrationally bound state in another potential. Similar structures have been seen in the cross section of HF$^+$~\cite{roos09}. Above the dissociation energy of the ion (around 2.66 eV) there are no bound vibrational levels of the Rydberg states and this explains the lack of structures at higher energies. Furthermore, the dynamics on the  highest $^2\Sigma^+$ resonant states is studied using uncoupled potentials. 

When the electronic states of $^2\Pi$ symmetry are diabatized, eight states are included, i.e., all five resonant states and three electronically bound states. The lowest $A^2\Pi$ state is not included in the process. 
\begin{figure}[h]
\rotatebox{0.0}{\includegraphics[width=0.9\columnwidth]{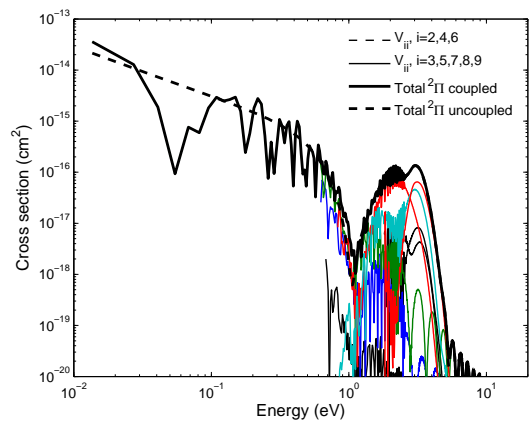}}
\caption{\label{fig8} Cross section calculated using the coupled states of BeH of $^2\Pi$ symmetry.
Also the cross section calculated using uncoupled resonant states is displayed.} 
\end{figure}
Again, the electronic couplings induce sharp oscillations in the cross sections, but they will not significantly effect the magnitude of the cross section.

The fact that the inclusion of the couplings do not increase the magnitude of the cross section at low energies may be understood by smooth onset of most of the calculated partial cross sections. Hence, it is the electron capture that
limits the energy components of the wave packets captured into the resonant states. Simply, there is no low-energy components initiated in the resonant states that can find a way to dissociation by inclusion of electronic couplings.

Fig.~\ref{fig8} shows the total cross section for dissociative recombination of BeH$^+$ calculated using both the uncoupled and coupled models. Both models produce a cross section with a similar magnitude, but oscillations are created when the electronic couplings are included.  
\begin{figure}[h]
\rotatebox{0.0}{\includegraphics[width=0.9\columnwidth]{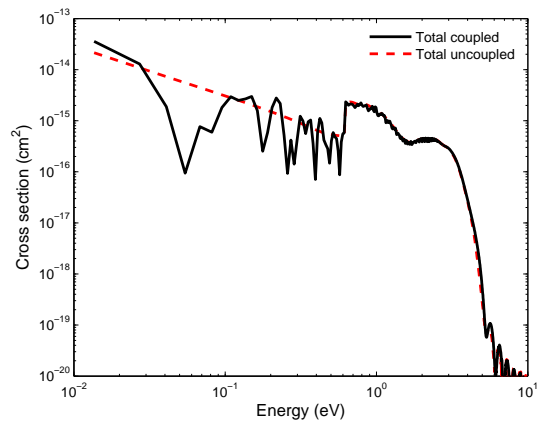}}
\caption{\label{fig9} Total cross section for dissociative recombination of BeH$^+$, calculated using coupled and uncoupled potentials.} 
\end{figure}
The cross section has a low-energy $1/E$ dependence and a magnitude of about $3\times10^{-14}$ cm$^2$ at $E=0.01$ eV.
This is a typical magnitude for a dissociative recombination cross section of a diatomic molecule~\cite{alkhalili03}.
Furthermore, the cross section shows a high-energy peak centered around 1 eV. Similar high-energy peaks have been seen in the cross section for dissociative recombination for several ionic species~\cite{alkhalili03}. The sharp threshold around 0.63 eV in the total cross section comes from the $^2\Delta$ resonant state.

\subsection{Effects of vibration}
\begin{figure}[h]
\rotatebox{0.0}{\includegraphics[width=0.9\columnwidth]{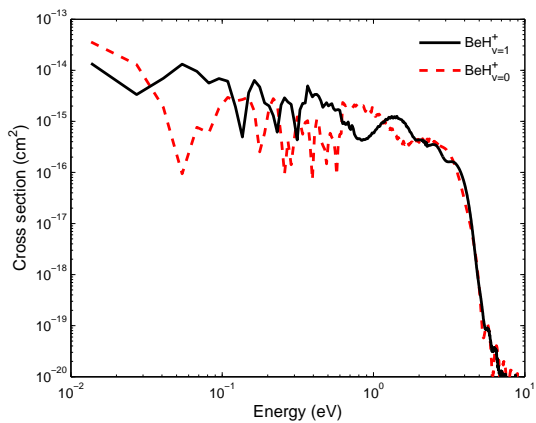}}
\caption{\label{fig10} Total cross section for dissociative recombination of BeH$^+$ in the $v=1$ and $v=1$ vibrational states.} 
\end{figure}

The cross section for dissociative recombination with BeH$^+$ in the $v=1$ vibrational level is calculated by using the $\chi_{v=1}$ vibrational wave function in the initial condition of the wave packets [eq. (\ref{eq:init})]. Furthermore, the energy-scale is shifted to account for the shift in energy of the target ion. The wave packets are then propagated on the coupled potentials. In Fig.~\ref{fig10}, we compare the total cross section for
dissociative recombination of BeH$^+$ in $v=0$ and $v=1$. 
It can be seen that the shift in the energy scale causes shifts of the resonant structures as well as the high energy peak towards smaller energies. Furthermore, the cross section at low collision energies is increased and this can be understood
by larger capture probability into the low lying resonant state for the vibrationally excited ion.

\subsection{Effects from isotopic substitution}
\begin{figure}[h]
\rotatebox{0.0}{\includegraphics[width=0.9\columnwidth]{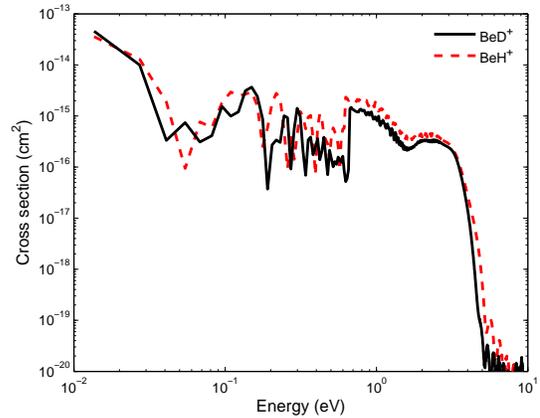}}
\caption{\label{fig11} Total cross section for dissociative recombination of BeD$^+$ and BeH$^+$ in their lowest $v=0$ vibrational states.} 
\end{figure}

The cross section for dissociative recombination with BeD$^+$ in the $v=0$ vibrational state is 
calculated by using the same set of potentials, widths and couplings, but changing
the reduced mass to the mass of BeD. Both when the vibrational wave function of the target ion is calculated 
as well as in the wave packet propagation the mass has to be changed. 
Again, the coupled potentials are used for calculating the cross section.
The resulting cross section is displayed in Fig.~\ref{fig11} and compared with the corresponding cross section for BeH$^+$.
The cross section of the two isotopologues are similar in shape and magnitude. They both show oscilliations induced by the electronic couplings as well as the high-energy peak around 1 eV. The magnitude of the high-energy peak is smaller for the heavier isotopologue and similar behaviours have been seen
for the high-energy peak oberserved in for example dissociative recombination with H$_3^+$~\cite{larsson97}.
The cross section for electron recombination with BeD$^+$ in the vibrationally excited $v=1$ state is calculated and similar to BeH$^+$, the positions of the structures in the cross section as well as the high energy peak are shifted in the energy and the magnitude of the low-energy cross section is increased.

\subsection{Analytical forms for the cross sections}
To simplify the use of the calculated cross sections in modelling of the fusion plasmas, the cross sections have been fitted to analytical forms. The cross sections calculated using uncoupled potentials have been used for the fitting. These cross sections show a smooth energy dependence at collision energies below about 0.3 eV that easily can be fitted to functions of the form
\begin{equation}
\sigma(E)=\frac{\sigma_0}{E^b}
\end{equation}
In table~\ref{tab:fit}, the values of the parameters $\sigma_0$ and $b$ are given for the fits of the calculated cross sections for dissociative recombination of
BeH$^+$ and BeD$^+$ in the $v=0$ and $v=1$ vibrational levels.
 
\begin{table}
\caption{\label{tab:fit} Parameters for the fitted form of the dissociative recombination cross sections at low collision energies.}
\begin{ruledtabular}
\begin{tabular}{cccc}
Ion &vibrational level & $\sigma_0$ ($10^{-16}$ cm$^2$$\times$eV$^b$) & $b$\\
\hline
\\
BeH$^+$& $v=0$ & $3.10$  & $0.99$ \\
BeH$^+$& $v=1$ & $6.90$  & $0.93$ \\
BeD$^+$& $v=0$ & $3.32$  & $1.02$ \\
BeD$^+$& $v=1$ & $4.72$  & $1.05$ \\
\end{tabular}
\end{ruledtabular}
\end{table}
\section{Conclusions}  
For the first time, the cross section for dissociative recombination of BeH$^+$ has been calculated. This cross section is important in the modeling of the fusion plasmas. However, the toxicity of BeH$^+$ makes measurements on the reaction extremely difficult. In the present study, structure and electron scattering calculations are combined to obtain qausidiabatic potentials of BeH. Using an optimization procedure, the electronic couplings between the resonant states are estimated.
The dynamics of the reaction is then explored using a wave packet technique. The reaction is studied assuming both uncoupled and coupled potentials. The inclusion of the couplings cause oscillations in the cross section and will not significantly affect the magnitude of the cross section. The cross section shows a low-energy $1/E$ dependence followed by a high-energy peak
centered around 1 eV. The cross section for dissociative recombination with vibrationally excited $v=1$ ions is larger at low collision energies than the corresponding cross section for vibrationally relaxed ions. The cross section for the heavier isotopologue BeD$^+$ is similar to the cross section for BeH$^+$, except for the high-energy peak that becomes smaller in magnitude. 

\begin{acknowledgments}
We are thankful for data of calculated potential curves send
to us by Jose Sanchez-Marin.
\AA. L. acknowledges support from The Swedish Research Council
and A. E. O. acknowledges support from the National Science 
Foundation, Grant No. PHY-05-55401. M. L. participated in the
IAEA Coordinated Resreach Project ``Atomic and Molecular Data for
Plasma Modelling".
\end{acknowledgments}

\end{document}